\documentclass[reprint,superscriptaddress,nofootinbib,amsmath,amssymb,aps]{revtex4-1}

\usepackage{graphicx}% Include figure files
\usepackage{dcolumn}% Align table columns on decimal point
\usepackage{bm}% bold math
\usepackage{color}

\def\prd{Phys. Rev. D}
\def\prl{Phys. Rev. Lett.}

\def\nat{Nature}

\begin{document}

\preprint{APS/123-QED}

\title{Exciting modes due to the aberration of gravitational waves:\\ 
Measurability for extreme-mass-ratio inspirals}

\author{Alejandro Torres-Orjuela}
\affiliation{Astronomy Department, School of Physics, Peking University, 100871 Beijing, China}
\affiliation{Kavli Institute for Astronomy and Astrophysics at Peking University, 100871 Beijing, China}

\author{Pau Amaro Seoane}
\affiliation{Universitat Polit{\`e}cnica de Val{\`e}ncia, 46022 Val{\`e}ncia, Spain}
\affiliation{Deutsches Elektronen Synchrotron DESY, Platanenallee 6, 15738 Zeuthen, Germany}
\affiliation{Kavli Institute for Astronomy and Astrophysics at Peking University, 100871 Beijing, China}
\affiliation{Institute of Applied Mathematics, Academy of Mathematics and Systems Science, Chinese Academy of Sciences, Beijing 100190, China}
\affiliation{Zentrum f{\"u}r Astronomie und Astrophysik, TU Berlin, Hardenbergstra{\ss}e 36, 10623 Berlin, Germany}

\author{Zeyuan Xuan}
\affiliation{Physics Department, School of Physics, Peking University, 100871 Beijing, China}

\author{Alvin J. K. Chua}
\affiliation{Jet Propulsion Laboratory, California Institute of Technology, Pasadena, CA 91109, USA}
\affiliation{Theoretical Astrophysics Group, California Institute of Technology, Pasadena, CA 91125, USA}

\author{María J. B. Rosell}
\affiliation{Theory Group, Department of Physics, University of Texas at Austin, Austin, TX 78712, USA}

\author{Xian Chen}
\email{Corresponding author: xian.chen@pku.edu.cn}
\affiliation{Astronomy Department, School of Physics, Peking University, 100871 Beijing, China}
\affiliation{Kavli Institute for Astronomy and Astrophysics at Peking University, 100871 Beijing, China}

\date{\today}

\begin{abstract}
Gravitational waves from a source moving relative to us can suffer from special-relativistic effects such as aberration.
The required velocities for these to be significant are on the order of $1000\,\textrm{km\,s}^{-1}$. This value corresponds to the velocity dispersion that one finds in clusters of galaxies. Hence, we expect a large number of gravitational-wave sources to have such effects imprinted in their signals.
In particular, the signal from a moving source will have its higher modes excited, i.e., $(3,3)$ and beyond. We derive expressions describing this effect, and study its measurability for the specific case of a circular, non-spinning extreme-mass-ratio inspiral.
We find that the excitation of higher modes by a peculiar velocity of $1000\,\textrm{km\,s}^{-1}$ is detectable for such inspirals with signal-to-noise ratios of $\gtrsim20$. Using a Fisher matrix analysis, we show that the velocity of the source can be measured to a precision of just a few percent for a signal-to-noise ratio of 100.
If the motion of the source is ignored parameter estimates could be biased, e.g., the estimated masses of the components through a Doppler shift. Conversely, by including this effect in waveform models, we could measure the velocity dispersion of clusters of galaxies at distances inaccessible to light.
\end{abstract}

                             % Classification Scheme.
                              %display desired
\maketitle

\textit{\textbf{Introduction} ---} In two recent papers \citep{Torres-OrjuelaEtAl2019,Torres-OrjuelaEtAl2020} by some of the present authors, we investigated several relativistic effects due to a moving source of gravitational waves (GWs). These effects share similarities with the beaming and aberration of light, but differ significantly in fundamental properties and distort the interpretation of the signal. We also addressed in \cite{Torres-OrjuelaEtAl2019} the effect of source motion on the wave polarization, and showed that this will lead to a rotation that can affect the amplitude of the detector response. 

The effect of a time-dependent motion on the modes of GWs has been studied in the context of gravitational recoil by different authors. In \cite{2016PhRvD..93h4031B}, the authors studied the effect that the kick imprints on numerical relativistic waveforms and presented a scheme to extract this information from the waveforms. Further, \cite{calderon-bustillo_clark_2018} showed that gravitational kicks could be detected with high accuracy by considering the anisotropic emission of GWs encoded by higher modes. The effects of a constant source motion on the modes of gravitational waves has also been studied in previous works. In particular, \cite{2008PhRvD..78d4024G} addressed the effect of motion (velocity) on the modes, imposing the restriction that the velocity is non-relativistic and parallel to the line-of-sight, and focused on modes with $\ell=2$. Recently, \cite{2019PhRvD.100l4010W} also looked into the role that motion can have on
numerical relativity waveforms as a numerical effect, and how to correct this.

In contrast to the case of kicks that induce a time-dependent motion, so far the effect of the modes on the detection of GWs for a constant velocity has not being considered in LIGO/Virgo detections~\cite{GWTC1,GWTC2}. In fact, because of Schutz's seminal work from 1986~\cite{schutz_1986} it is widely believed that a constant motion of the source only induces a constant redshift which is degenerate with the total mass of the source. However, in Schutz's work only the dominant quadrupole mode is considered and subsequent works (including a recent paper by some of the present authors~\cite{torres-orjuela_chen_2020b}) have shown that this picture changes when considering additional modes.

For the effect of the motion to alter the observation of the gravitational wave, it is crucial that the source is moving relative to us at a relatively large speed, of at least a few hundreds of kilometers per second. It is important to note that what matters is the relative velocity of the center-of-mass (CoM) of the source relative to us and not how fast the binary components are moving relative to the CoM or us. Moreover, we are talking about the peculiar velocities of the galaxies host to the gravitational wave sources, and \textit{not} the cosmological expansion, which is already taken into account in current detections. This effect will be present in all sources of gravitational waves, but it will be most relevant for LISA sources such as extreme-mass-ratio inspirals (EMRIs), because of the long signal duration and high eccentricity \cite{amaro-seoane_2018a}.

\textit{\textbf{Peculiar velocity of galaxies} ---} Measurements of the bulk average motion we experience with respect to matter outside our galaxy have shown that this motion grows as a function of distance to us, going from about $300\,\textrm{km\,s}^{-1}$ at 300 Mpc \citep{6dF2015} to some $1700\,\textrm{km\,s}^{-1}$ at 6000 Mpc \citep{ColinEtAl2017}. This is the radius within which most of the LIGO/Virgo binaries will be located. This means about 40\% of the measured binaries will have bulk velocities above $1000\,\textrm{km\,s}^{-1}$ relative to us.

\begin{figure}
\resizebox{\hsize}{!}
          {\includegraphics[scale=1,clip]{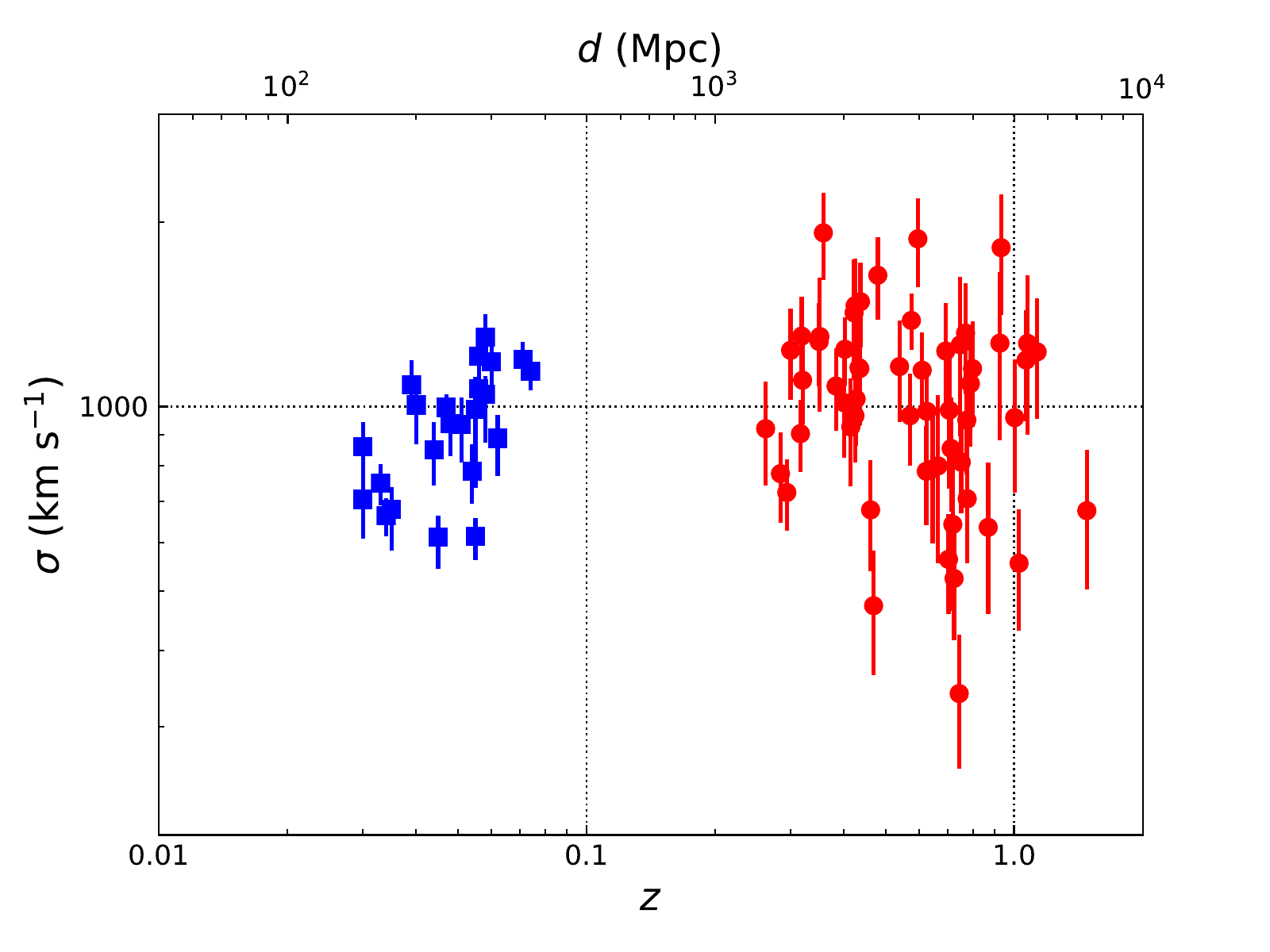}}
\caption
   {
    Velocity dispersion of galaxy clusters as a function of the redshift, $z$, (bottom X-axis) and the distance, $d$, (upper X-axis) from the data of \cite{GirardiEtAl1996} (blue squares) and \cite{RuelEtAl2014} (red circles). We add the error bars for both the velocity dispersion and the distance, although the latter does not show up because it is smaller in size than the symbols.
   }
\label{fig.VelDis}
\end{figure}

In addition to the bulk dipole, about 20\% of galaxies are in rich galaxy clusters \citep{Bahcall1988}, orbiting around the CoM of the cluster with velocity dispersions of $1000\,\textrm{km\,s}^{-1}$ \citep{Zwicky1933}. In Fig.~\ref{fig.VelDis}, we show the velocity dispersion, $\sigma$, of several galaxy clusters (a point is a large collection of galaxies, and each galaxy harbors some $10^{11}$ stars) as a function of redshift, $z$, we have made from the data from \cite{GirardiEtAl1996,RuelEtAl2014}. Rich galaxy clusters contain a higher number of higher mass, elliptical galaxies. This means that the percentage of gravitational-wave sources will be even higher, because the host galaxies are larger, they contain more stars and compact objects.  

Overall, given that the two motions we have described should be uncorrelated, one would expect that about 60\% of all gravitational wave sources move with a velocity of $1000\,\textrm{km\,s}^{-1}$ or more relative to us. As we will show in the next section such high velocities induce an excitation of additional modes for all sources of GWs. However, we focus in the subsequent section on the impact for EMRIs, because they have relatively high signal-to-noise ratios , which makes it easier to detect the velocity.

\textit{\textbf{Excitation of modes} ---} As discussed in Ref~\cite{Torres-OrjuelaEtAl2019}, a motion of the CoM of a GWs source will affect the GWs emitted through aberration. For light it is well known that aberration will affect the luminosity of the source~\cite{johnson_teller_1982} and that this can be used to measure a relative velocity between the observer and the source (see, e.g.,~\cite{ColinEtAl2017}). For GWs we do not measure the luminosity of the source but the wave's $+$- and $\times$-polarization, $h_{+,\times}$. These two polarizations can be combined to form the complex amplitude, $H(\theta,\phi) := h_+(\theta,\phi) - ih_\times(\theta,\phi)$, which is conveniently decomposed into its spherical modes, $H^{\ell,m}$, using spin-2 spherical harmonics, $_{-2}Y^{\ell,m}(\theta,\phi)$,~\cite{thorne_1980,ruiz_alcubierre_2008,goldberg_macfarlane_1967}

\begin{equation}\label{eq:comamp}
    H(\theta,\phi) = \sum_{\ell=2}^\infty\sum_{m=-\ell}^\ell H^{\ell,m}\,_{-2}Y^{\ell,m}(\theta,\phi),
\end{equation}

\noindent
where

\begin{equation}\label{eq:modes}
    H^{\ell,m} := \int H(\theta,\phi)_{-2}\bar{Y}^{\ell,m}(\theta,\phi) \text{d}\Omega,
\end{equation}

\noindent
$\theta$ and $\phi$ are the polar and azimuthal angles of the source, and $_{-2}\bar{Y}^{\ell,m}(\theta,\phi)$ denotes the complex conjugate of $_{-2}Y^{\ell,m}(\theta,\phi)$. The modes of the wave are related to the spherical properties of the source or simply speaking to its spherical ``shape''. If the source is now moving, aberration affects the shape of the source seen by a distant observer and thus the observer sees modes that differ from those of a source at rest. This difference in the modes can then be used to detect a motion of the source relative to the observer.

Here we derive briefly how the modes of a GWs source are affected through aberration up to linear order in the velocity of the source relative to the observer, $\boldsymbol{v} = (v_x,v_y,v_z)$. A more extensive analysis and derivation of the effect of the velocity on the modes of GWs has been published by some of the present authors in Ref.~\cite{torres-orjuela_chen_2020b}. We use a coordinate system for which the $z$-coordinate lies along the Newtonian angular momentum of the source, denote the polar angle measured relative to the $z$-coordinate by $\theta$ and the azimuthal angle measured from the $x$-coordinate by $\phi$. Further, we will mark the quantities measured in the rest frame of the distant observer that sees the source moving by a prime.

Due to aberration a distant observer sees a ray pointing in the direction $\boldsymbol{m}$ in the source frame to point in his own frame in the direction~\cite{Torres-OrjuelaEtAl2020}

\begin{equation}
    \boldsymbol{m}' = \boldsymbol{m} + \langle\boldsymbol{m},\boldsymbol{v}\rangle\boldsymbol{m} - \boldsymbol{v},
\end{equation}

\noindent
where $\langle\cdot,\cdot\rangle$ denotes the three-dimensional Euclidean scalar product and we only have considered contributions up to linear order in the velocity. When expressed in spherical coordinates that means that a ray pointing in the $(\theta,\phi)$ direction in the source frame points in the $(\theta',\phi')$ direction in the observer frame, where

\begin{align}
    \cos(\theta') =& (1 + \langle\boldsymbol{m},\boldsymbol{v}\rangle)\cos(\theta) - v_z, \\
    \tan(\phi') = & \tan(\phi) + \frac{1}{\sin(\theta)\cos(\phi)}\big[\tan(\phi)v_x - v_y\big].
\end{align}

The complex amplitude is a scalar function of the spherical coordinates. Therefore, aberration causes that a moving observer sees the same complex amplitude but relative to the `new spherical coordinates' $(\theta',\phi')$. Therefore, we get

\begin{equation}
    H'(\theta',\phi') \Leftrightarrow H(\theta,\phi),
\end{equation}

\noindent
where by the equivalence symbol we mean that $H'$ takes relative to the coordinates $(\theta',\phi')$ the same values as $H$ takes relative to the coordinates $(\theta,\phi)$. Expanding $H'(\theta',\phi')$ to linear order in the velocity of the source, we get

\begin{align}
H'(\theta',\phi') & = H(\theta,\phi) \nonumber \\
                & + \frac{1}{\sin(\theta)}\big[(\partial_{\theta}H(\theta,\phi)) (v_z- \langle\boldsymbol{m},\boldsymbol{v}\rangle\cos(\theta)) \nonumber \\
                & + (\partial_{\phi}H(\theta,\phi))(v_x\sin(\phi) - v_y\cos(\phi))\big],
\end{align}

\noindent
and because the right hand side of this equation is only a function of $\theta$ and $\phi$, we can treat henceforth $H'$ as a function of these two.

Using Eq.~(\ref{eq:comamp}) and the differential properties of the spin-2 spherical
harmonics~\cite{ruiz_alcubierre_2008}, we find

\begin{align}
    H'(\theta,\phi) =& \sum_{\ell = 2}^\infty\sum_{m = -\ell}^{\ell} \left[1 + f^{\ell,m}(\theta,\phi)\right] \\
    \nonumber & H^{\ell,m}\,_{-2}Y^{\ell,m}(\theta,\phi),
\end{align}

\noindent
where $H^{\ell,m}$ are the modes in the rest frame of the source and we have defined

\begin{align}
& f^{\ell,m}(\theta,\phi)  := \frac{1}{\sin(\theta)} \Bigg[i\,m(v_z - \langle\boldsymbol{m},\boldsymbol{v}\rangle\cos(\theta)) \nonumber \\
                        & + \frac12\Big(\sqrt{(\ell-m)(\ell+m+1)} - \sqrt{(\ell+m)(\ell-m+1)}\Big) \nonumber \\
                        & ~~\Big(v_x\sin(\phi) - v_y\cos(\phi)\Big)\Bigg].
\end{align}

If we now decompose $H'(\theta,\phi)$ in the same way as in Eq.~(\ref{eq:modes}), we find for the modes of the moving source

\begin{align}\label{eq:ampcha}
& H'^{\ell,m} = H^{\ell,m} \nonumber \\
                         & + (-1)^m i \bigg[\sum_{\ell_0 = \textrm{max}\{2,\,|m|\}}^\infty C_0(\ell_0,\ell,m) H^{\ell_0,m} \nonumber \\
                         & + \sum_{\ell_+ = \textrm{max}\{2,\,|m+1|\}}^\infty C_+(\ell_+,\ell,m) H^{\ell_+,m+1} \nonumber \\
                         &+ \sum_{\ell_- = \textrm{max}\{2,\,|m-1|\}}^\infty C_-(\ell_-,\ell,m) H^{\ell_-,m-1}\bigg],
\end{align}

\noindent
where $C_0$, $C_+$ and $C_-$ are coefficients proportional to the magnitude of the velocity. This equation tells us that the amplitudes of the different modes are changed due to the motion of the source. The polarizations of the source are the real and minus imaginary part of a combination of these modes and hence altered by their change. Therefore, the amplitude of the two polarizations can either increase or diminish, depending on the particular combination of the modes, the velocity and the direction from which the source is observed. Moreover, because the different modes enter with a different contribution to the phase of the wave~\cite{ArunEtAl2009}, it also will lead to a shift of the total phase of the wave. Since the amplitude of the modes is time-dependent both effects on the polarizations are time-dependent, too, even for a constant velocity~\cite{torres-orjuela_chen_2020b}.

\textit{\textbf{Measurability of the effect} ---} In the previous section, we show that a constant motion of the source leads to an excitation of the modes so that the GW signal from such a source, $h'$, differs from the signal of a source at rest, $h$. This difference in the two signals can in principle be distinguished if it is big enough. To understand the measurability of this effect, we have performed an approximate data-analysis study in the context of LISA.

We first introduce some standard definitions in gravitational-wave signal analysis. We treat the waveforms as vectors in a Hilbert space \cite{Helstrom68}, which allows us to define the noise-weighted inner product

\begin{equation}\label{eq:inpro}
    \left<h'\left|h\right.\right> := 2\int_{0}^{\infty}{df}\,\frac{  \tilde{h'}(f)\tilde{h}(f)^{*} + \tilde{h'}(f)^{*}\tilde{h}(f)}{S_{n}(f)},
\end{equation}

\noindent
where $\tilde{h}(f)$ is the Fourier transform of the time domain waveform $h(t)$. In this expression, $S_{n}(f)$ is the one-sided noise power spectral density of LISA \cite{Thorne87,Finn92,2016PhRvD..93b4003K}.

In the long-wavelength approximation, LISA data can be post-processed to give two noise-orthogonal channels $d_\mathrm{I,II} = h_\mathrm{I,II}+n_\mathrm{I,II}$, where $h_\mathrm{I,II}$ are the projections of $h$ onto each channel, and $n_\mathrm{I,II}$ are the noise realizations in each channel. For this simple analysis, we will find it convenient to consider only the GW polarizations $h_{+,\times}$ at the Solar-System barycenter instead of $h_\mathrm{I,II}$ (these are related through a time-evolving and Doppler-shifted linear combination that encodes the detector motion). If we adopt the assumption that LISA noise is stationary, Gaussian, and characterized by $S_{n}(f)$, the optimal signal-to-noise ratio (SNR) when filtering the data against $h$ is given by

\begin{equation}
    \rho = \sqrt{\langle h_+ | h_+ \rangle + \langle h_\times | h_\times \rangle}.
\end{equation}

Given two different waveforms $h'$ (the observed one) and $h$ (the putative one), the usual way to quantify their difference is to evaluate their overlap, or match:

\begin{equation}\label{eq:match}
    \mathrm{M} := \frac{\left\langle h' | h \right\rangle}{\sqrt{\left\langle h' | h' \right\rangle\left\langle h | h \right\rangle}},
\end{equation}

\noindent
so that $\mathrm{M}=1$ when there is a perfect match between the putative and the observed waveform.

While the match shows the similarity between two waveforms, whether we can distinguish these two waveforms depends additionally on how strong the signal is. One way of estimating this is to require that the shift in recovered parameters when using a waveform $h$ to measure a signal $h'$ exceeds the expected statistical error due to detector noise (when using $h'$ to measure $h'$). Through analyses such as in Ref.~\cite{CutlerVallisneri2007}, we can derive a rough rule-of-thumb criterion for the two waveforms to be distinguishable: $\rho > \sqrt{D/(2(1-\mathrm{M}))}$. Here, $D$ represents the number of parameters to be measured in the analysis, which for a typical LISA source can be set to $D=10$.

If the two waveforms can be distinguished, another question is how precisely the parameters of the source, $\boldsymbol{\lambda}$, can be extracted from the signal. This question can be addressed using a Fisher matrix analysis~\citep{coe_2009}, which provides a linearized estimate for the measurement errors that asymptotes to the true errors in the high-SNR limit. The Fisher matrix can be calculated as

\begin{equation}
    F_{ij} := \left\langle\frac{\partial h(\boldsymbol{\lambda})}{\partial \lambda_i},\frac{\partial h(\boldsymbol{\lambda})}{\partial \lambda_j}\right\rangle.
\end{equation}

\noindent
The inverse of the Fisher matrix, $C = F^{-1}$, approximates the sample covariance matrix of the Bayesian posterior distribution for the parameters given the observed signal.

A more detailed study of the detection of a CoM velocity and its effect on parameter estimation would take into account parameter correlations in realistic waveforms (e.g., \cite{ChuaEtAl2020}), and involve analyses such as in Ref.~\cite{CutlerVallisneri2007} or full posterior sampling. However, in this work we only intend to show that the velocity can, in principle, be detected and measured. Thus we stick to simple analytic waveforms and the analysis introduced above.

As an example, we examine an EMRI of two non-spinning black holes (BHs) on a \textit{circular} orbit in the LISA band for simplicity. We note, however, that EMRIs form at very high eccentricities \citep{amaro-seoane_2018a}, and any residual eccentricity very likely will enhance the effect we describe in this work. A circular orbit represents the most conservative scenario, since higher modes are more prominent for eccentric orbits \cite{peters_mathews_1963}.

We construct the putative EMRI waveform, $h(t)$, following the post-Newtonian prescription of Ref.~\cite{ArunEtAl2009}. This allows us to easily generate waveforms containing the most important modes up to $(\ell,|m|) = (5,5)$. We analyze the representative case of a stellar mass BH ($m_2 = 10\,M_\odot$) inspiraling into a super-massive BH ($m_1 = 10^6\,M_\odot$), where the signal is observed for the final two years before plunge (see Ref.~\cite{BabakEtAl2017} for a similar case). For the observed waveform, $h'$, we use the same conditions but distort the modes according to Eq.~(\ref{eq:ampcha}), where we assume a CoM velocity of 1000\,km/s pointing along the Newtonian angular momentum of the source. Note also that we set the masses of the BHs to be the same in the observer frame, to correct for the Doppler effect.

\begin{figure}
\resizebox{\hsize}{!}
          {\includegraphics[scale=1,clip]{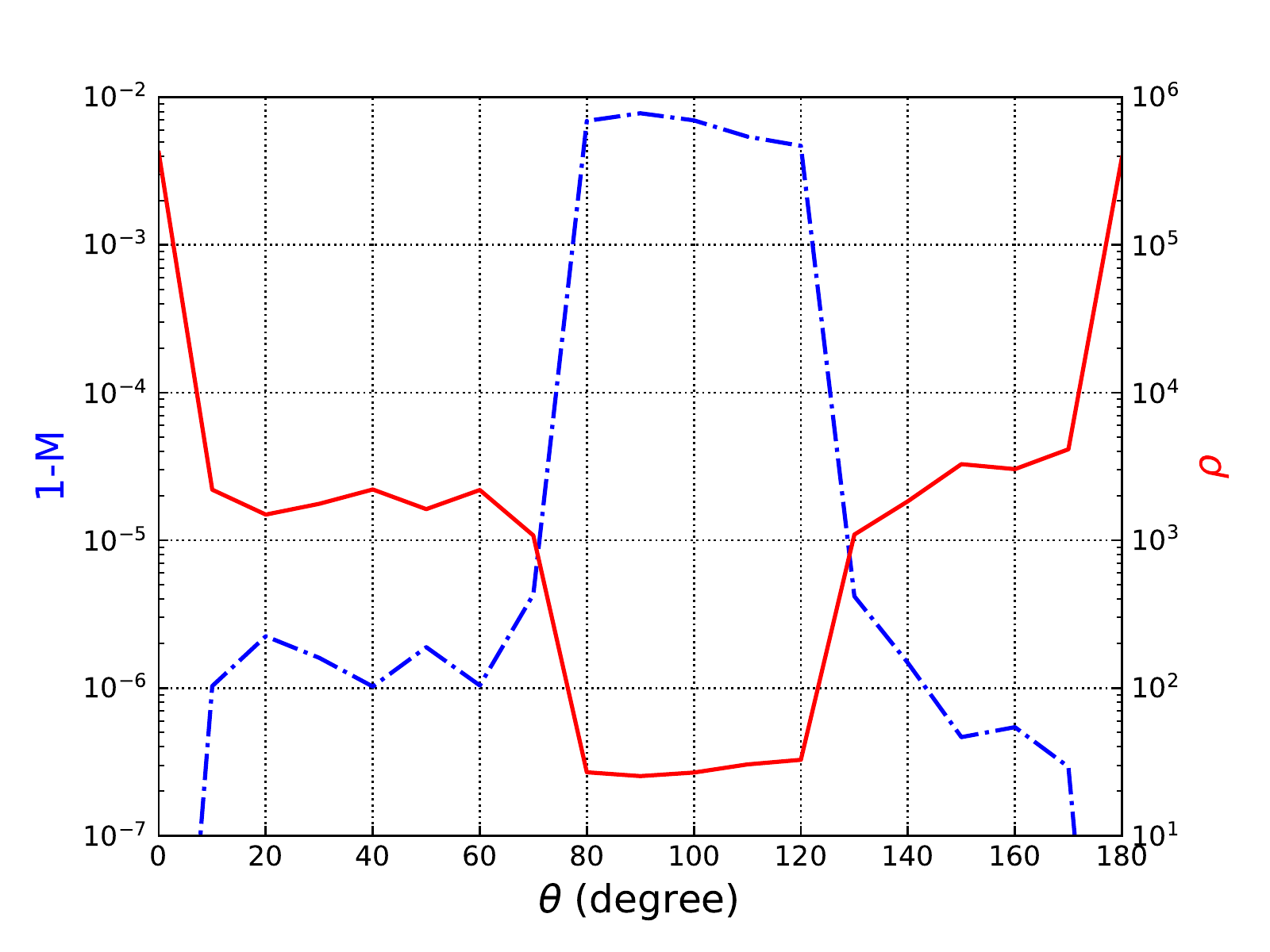}}
\caption
   {
    Mismatch, $1-\textrm{M}$, between the putative waveform of an EMRI at rest and the observed waveform of an EMRI moving with a typical CoM velocity of 1000\,km/s (dotted blue line/left Y-axis) and the SNR, $\rho$, required to detect this mismatch (red continuous line/right Y-axis). Both are shown for different viewing angles of the source, $\theta$, where $\theta = 0^\circ$ corresponds to the source seen face on.
    }
\label{fig.MMSNR} 
\end{figure}

In Fig.~\ref{fig.MMSNR} we show the mismatch, $1-M$, between the two waveforms and the SNR that would be required to detect the respective mismatch (using the above rule-of-thumb criterion) for different viewing angles of the source. The highest values of the mismatch of several times $10^{-3}$ occur for viewing angles of $80^\circ$--$120^\circ$. Such mismatches could be resolved for a relatively low SNR of 20 to 30. For higher SNRs of a few 100, we would be able to resolve smaller mismatches down to $10^{-5}$, extending the range of viewing viewing angles between $70^\circ$ and $130^\circ$. These numbers all fall within the plausible range of EMRI SNRs ($\rho\lesssim10^3$) for a variety of astrophysical models \cite{BabakEtAl2017}.

\begin{figure}
\resizebox{\hsize}{!}
          {\includegraphics[scale=1,clip]{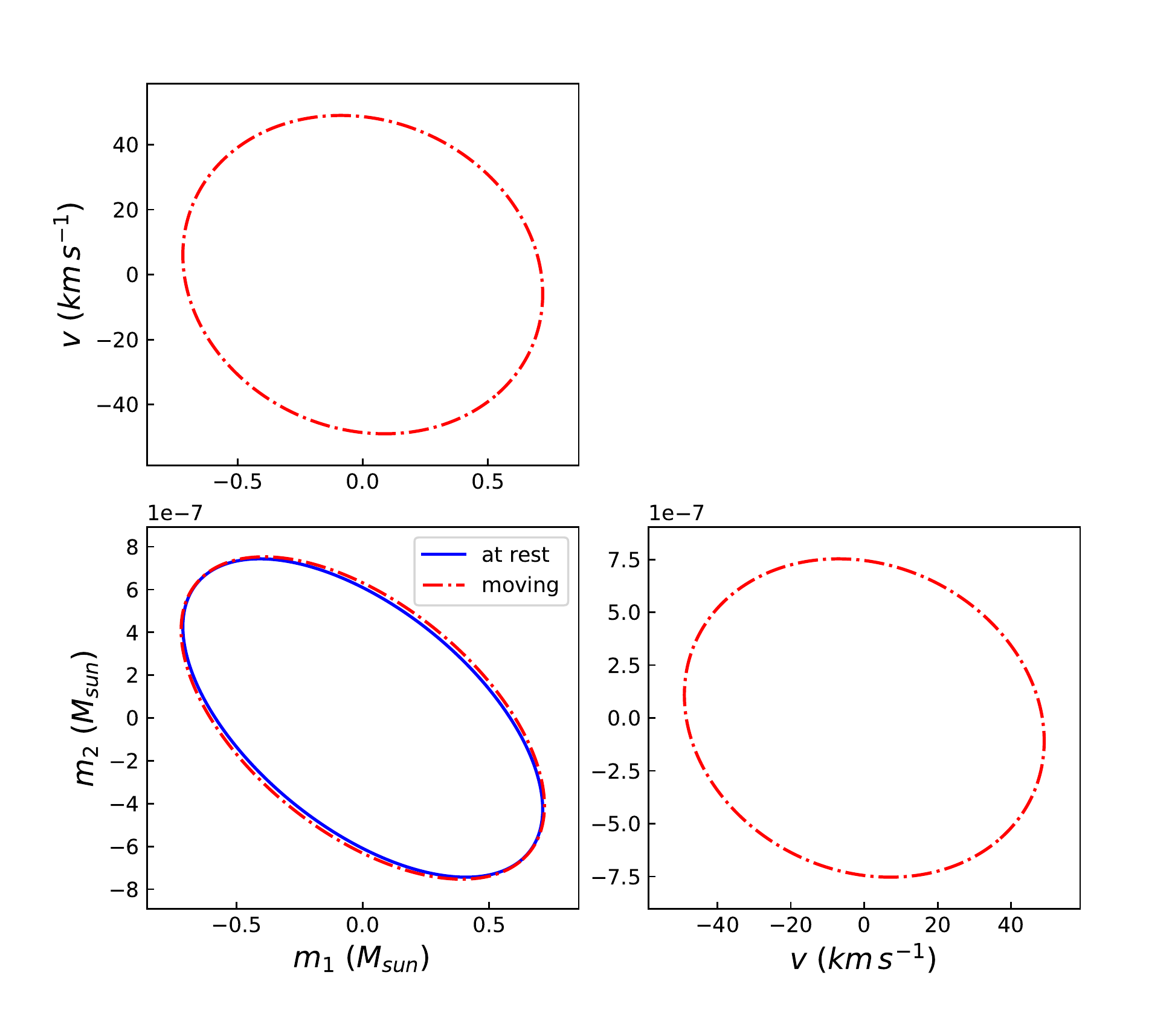}}
\caption
   {
    Confidence ellipses ($1\sigma$-level) for the measurement of the masses of the BHs, $m_{1,2}$, and the CoM velocity of the source in the case of a source at rest (solid blue line) and a moving source (dotted red line). Note that for the moving source the masses measured represent the (Doppler shifted) masses as seen in the observer frame.
    }
\label{fig.Covel} 
\end{figure}

We estimate how precisely the parameters of the sources can be extracted using a Fisher matrix analysis for a detection with an SNR of 100. In Fig.~\ref{fig.Covel}, we plot the confidence ellipses at the $1\sigma$-level for the masses of the components and the velocity of the source. The velocity of the source can be detected with a precision of about $40\,\textrm{km\,s}^{-1}$, corresponding to a few percent of the CoM velocity of the source. In Fig.~\ref{fig.VelDis}, where the velocities were measured using light, we see that the errors are of the order 10\,\%. For the velocity dispersion it is necessary to average over several sources and hence it is not surprising that the errors are bigger than for a single source, as we are considering here. However, this simple comparison indicates that the accuracy of the velocities measured with GWs is at least comparable to measurements with light.

From Fig.~\ref{fig.Covel} we, further, see that the two masses can be measured with an accuracy of around $10^{-6}$, independent of whether the source is at rest or moving. The CoM velocity is weakly correlated with the mass parameters, at least in the case of this simple circular model. Nevertheless, it should be noted that the accuracy refers to the masses as seen in the observer frame. For the moving source the masses in the source frame differ by a Doppler shift, which for a velocity of $1000\,\textrm{km\,s}^{-1}$ corresponds to a difference of $\sim 10^{-4}$. Therefore, an observer unaware of the Doppler shift would be estimating the masses with an error of around $10^{-4}$ and only when knowing the magnitude of the velocity can an accuracy of the order $10^{-6}$ be reached.

In the future, electro-magnetic (EM) counterparts to the moving EMRIs could serve to further calibrate and to study potential systematic errors of the technique presented in this work. Such EM counterparts could be detected as precursors in the X-ray band~\cite{mckernan_ford_2019} for EMRIs formed in the disks of active galactic nuclei~\cite{pan_yang_2021,pan_lyu_2021,barausse_rezzolla_2008}. Further, EM counterparts could be studied by comparing the velocity dispersion of a galaxy cluster measured with GWs to measurements with light~\cite{GirardiEtAl1996,RuelEtAl2014}. However, the viability of both methods might be restricted by our ability to resolve the location of EMRIs in order to locate their host systems with enough accuracy~\cite{lisa_2017}.

\textit{\textbf{Conclusions} ---} In this work, we have shown that a large number of host galaxies will have dispersion velocities of $\sim 1000\,\textrm{km\,s}^{-1}$. We have also derived general expressions for the excitation of higher modes in a moving source, which depend on the modes of the source in its own rest frame and on the CoM velocity of the source relative to the observer. This effect applies to all sources of gravitational waves, but in this work we focus, as an example, on its measurability for an important class of LISA source: an EMRI of a stellar-mass BH falling into a supermassive one (assuming that it is circular). We find that a peculiar velocity of $1000\,\textrm{km\,s}^{-1}$ could be detected for EMRIs with a SNR of more than 20 and in the case of a high but plausible SNR of around 100 the magnitude of the velocity could be measured with an accuracy of just a few percent. Our findings are also conservative, since we expect EMRIs to have residual eccentricity when they enter the LISA band, which will increase the presence of higher modes.

A detailed derivation of the excitation of the modes by a CoM velocity has been published by some of the present authors in Ref.~\cite{torres-orjuela_chen_2020b}. The investigation of data analysis implications are being analyzed in more detail, and will be presented soon elsewhere. We conclude by remarking that since the effect may be significant for EMRIs, its inclusion in waveform models can be used to measure peculiar motions of the host galaxies to distances which are inaccessible to light, and hence to obtain a mapping of galaxy cluster dispersions to redshifts as high as we can detect (circular) EMRIs with an SNR $\gtrsim\,20$. For eccentric EMRIs, the distances should be significantly larger.

\textit{\textbf{Acknowledgments} ---} ATO and XC acknowledge the support from the National Science Foundation of China, grants No.~11873022 and 11991053, and from the Strategic Priority Research Program of the Chinese Academy of Sciences, grants No.~XDB23040100 and XDB23010200. PAS acknowledges support from the Ram{\'o}n y Cajal Programme of the Ministry of Economy, Industry and Competitiveness of Spain, as well as the COST Action GWverse CA16104. This work was supported by the National Key R\&D Program of China (2016YFA0400702) and the National Science Foundation of China (11721303). AJKC is grateful to Christopher Moore for discussions on boosted gravitational waves, and acknowledges support from the NASA grant 18-LPS18-0027. MJBR has been supported by NSF Grant PHY-1912578. We thank Volker Springel, Massimo Dotti, Raffaella Schneider, Rosa Valiante, and Paolo Tozzi for discussions about galaxy clusters, and Enrico Barausse and K. G. Arun for discussions about the amplitudes on the subdominant modes.


\begin{thebibliography}{36}%
\makeatletter
\providecommand \@ifxundefined [1]{%
 \@ifx{#1\undefined}
}%
\providecommand \@ifnum [1]{%
 \ifnum #1\expandafter \@firstoftwo
 \else \expandafter \@secondoftwo
 \fi
}%
\providecommand \@ifx [1]{%
 \ifx #1\expandafter \@firstoftwo
 \else \expandafter \@secondoftwo
 \fi
}%
\providecommand \natexlab [1]{#1}%
\providecommand \enquote  [1]{``#1''}%
\providecommand \bibnamefont  [1]{#1}%
\providecommand \bibfnamefont [1]{#1}%
\providecommand \citenamefont [1]{#1}%
\providecommand \href@noop [0]{\@secondoftwo}%
\providecommand \href [0]{\begingroup \@sanitize@url \@href}%
\providecommand \@href[1]{\@@startlink{#1}\@@href}%
\providecommand \@@href[1]{\endgroup#1\@@endlink}%
\providecommand \@sanitize@url [0]{\catcode `\\12\catcode `\$12\catcode
  `\&12\catcode `\#12\catcode `\^12\catcode `\_12\catcode `\%12\relax}%
\providecommand \@@startlink[1]{}%
\providecommand \@@endlink[0]{}%
\providecommand \url  [0]{\begingroup\@sanitize@url \@url }%
\providecommand \@url [1]{\endgroup\@href {#1}{\urlprefix }}%
\providecommand \urlprefix  [0]{URL }%
\providecommand \Eprint [0]{\href }%
\providecommand \doibase [0]{http://dx.doi.org/}%
\providecommand \selectlanguage [0]{\@gobble}%
\providecommand \bibinfo  [0]{\@secondoftwo}%
\providecommand \bibfield  [0]{\@secondoftwo}%
\providecommand \translation [1]{[#1]}%
\providecommand \BibitemOpen [0]{}%
\providecommand \bibitemStop [0]{}%
\providecommand \bibitemNoStop [0]{.\EOS\space}%
\providecommand \EOS [0]{\spacefactor3000\relax}%
\providecommand \BibitemShut  [1]{\csname bibitem#1\endcsname}%
\let\auto@bib@innerbib\@empty
\bibitem [{\citenamefont {{Torres-Orjuela}}\ \emph {et~al.}(2019)\citenamefont
  {{Torres-Orjuela}}, \citenamefont {{Chen}}, \citenamefont {{Cao}},
  \citenamefont {{Amaro-Seoane}},\ and\ \citenamefont
  {{Peng}}}]{Torres-OrjuelaEtAl2019}%
  \BibitemOpen
  \bibfield  {author} {\bibinfo {author} {\bibfnamefont {A.}~\bibnamefont
  {{Torres-Orjuela}}}, \bibinfo {author} {\bibfnamefont {X.}~\bibnamefont
  {{Chen}}}, \bibinfo {author} {\bibfnamefont {Z.}~\bibnamefont {{Cao}}},
  \bibinfo {author} {\bibfnamefont {P.}~\bibnamefont {{Amaro-Seoane}}}, \ and\
  \bibinfo {author} {\bibfnamefont {P.}~\bibnamefont {{Peng}}},\ }\href
  {\doibase 10.1103/PhysRevD.100.063012} {\bibfield  {journal} {\bibinfo
  {journal} {Ph. Rv. D}\ }\textbf {\bibinfo {volume} {100}},\ \bibinfo {eid}
  {063012} (\bibinfo {year} {2019})},\ \Eprint
  {http://arxiv.org/abs/1806.09857} {arXiv:1806.09857 [astro-ph.HE]}
  \BibitemShut {NoStop}%
\bibitem [{\citenamefont {{Torres-Orjuela}}\ \emph
  {et~al.}(2020{\natexlab{a}})\citenamefont {{Torres-Orjuela}}, \citenamefont
  {{Chen}},\ and\ \citenamefont {{Amaro-Seoane}}}]{Torres-OrjuelaEtAl2020}%
  \BibitemOpen
  \bibfield  {author} {\bibinfo {author} {\bibfnamefont {A.}~\bibnamefont
  {{Torres-Orjuela}}}, \bibinfo {author} {\bibfnamefont {X.}~\bibnamefont
  {{Chen}}}, \ and\ \bibinfo {author} {\bibfnamefont {P.}~\bibnamefont
  {{Amaro-Seoane}}},\ }\href {\doibase 10.1103/PhysRevD.101.083028} {\bibfield
  {journal} {\bibinfo  {journal} {Ph. Rv. D}\ }\textbf {\bibinfo {volume}
  {101}},\ \bibinfo {eid} {083028} (\bibinfo {year} {2020}{\natexlab{a}})},\
  \Eprint {http://arxiv.org/abs/2001.00721} {arXiv:2001.00721 [astro-ph.HE]}
  \BibitemShut {NoStop}%
\bibitem [{\citenamefont {{Boyle}}(2016)}]{2016PhRvD..93h4031B}%
  \BibitemOpen
  \bibfield  {author} {\bibinfo {author} {\bibfnamefont {M.}~\bibnamefont
  {{Boyle}}},\ }\href {\doibase 10.1103/PhysRevD.93.084031} {\bibfield
  {journal} {\bibinfo  {journal} {\prd}\ }\textbf {\bibinfo {volume} {93}},\
  \bibinfo {eid} {084031} (\bibinfo {year} {2016})},\ \Eprint
  {http://arxiv.org/abs/1509.00862} {arXiv:1509.00862 [gr-qc]} \BibitemShut
  {NoStop}%
\bibitem [{\citenamefont {{Calder{\'o}n Bustillo}}\ \emph
  {et~al.}(2018)\citenamefont {{Calder{\'o}n Bustillo}}, \citenamefont
  {{Clark}}, \citenamefont {{Laguna}},\ and\ \citenamefont
  {{Shoemaker}}}]{calderon-bustillo_clark_2018}%
  \BibitemOpen
  \bibfield  {author} {\bibinfo {author} {\bibfnamefont {J.}~\bibnamefont
  {{Calder{\'o}n Bustillo}}}, \bibinfo {author} {\bibfnamefont {J.~A.}\
  \bibnamefont {{Clark}}}, \bibinfo {author} {\bibfnamefont {P.}~\bibnamefont
  {{Laguna}}}, \ and\ \bibinfo {author} {\bibfnamefont {D.}~\bibnamefont
  {{Shoemaker}}},\ }\href {\doibase 10.1103/PhysRevLett.121.191102} {\bibfield
  {journal} {\bibinfo  {journal} {\prl}\ }\textbf {\bibinfo {volume} {121}},\
  \bibinfo {eid} {191102} (\bibinfo {year} {2018})},\ \Eprint
  {http://arxiv.org/abs/1806.11160} {arXiv:1806.11160 [gr-qc]} \BibitemShut
  {NoStop}%
\bibitem [{\citenamefont {{Gualtieri}}\ \emph {et~al.}(2008)\citenamefont
  {{Gualtieri}}, \citenamefont {{Berti}}, \citenamefont {{Cardoso}},\ and\
  \citenamefont {{Sperhake}}}]{2008PhRvD..78d4024G}%
  \BibitemOpen
  \bibfield  {author} {\bibinfo {author} {\bibfnamefont {L.}~\bibnamefont
  {{Gualtieri}}}, \bibinfo {author} {\bibfnamefont {E.}~\bibnamefont
  {{Berti}}}, \bibinfo {author} {\bibfnamefont {V.}~\bibnamefont {{Cardoso}}},
  \ and\ \bibinfo {author} {\bibfnamefont {U.}~\bibnamefont {{Sperhake}}},\
  }\href {\doibase 10.1103/PhysRevD.78.044024} {\bibfield  {journal} {\bibinfo
  {journal} {\prd}\ }\textbf {\bibinfo {volume} {78}},\ \bibinfo {eid} {044024}
  (\bibinfo {year} {2008})},\ \Eprint {http://arxiv.org/abs/0805.1017}
  {arXiv:0805.1017 [gr-qc]} \BibitemShut {NoStop}%
\bibitem [{\citenamefont {{Woodford}}\ \emph {et~al.}(2019)\citenamefont
  {{Woodford}}, \citenamefont {{Boyle}},\ and\ \citenamefont
  {{Pfeiffer}}}]{2019PhRvD.100l4010W}%
  \BibitemOpen
  \bibfield  {author} {\bibinfo {author} {\bibfnamefont {C.~J.}\ \bibnamefont
  {{Woodford}}}, \bibinfo {author} {\bibfnamefont {M.}~\bibnamefont {{Boyle}}},
  \ and\ \bibinfo {author} {\bibfnamefont {H.~P.}\ \bibnamefont {{Pfeiffer}}},\
  }\href {\doibase 10.1103/PhysRevD.100.124010} {\bibfield  {journal} {\bibinfo
   {journal} {\prd}\ }\textbf {\bibinfo {volume} {100}},\ \bibinfo {eid}
  {124010} (\bibinfo {year} {2019})},\ \Eprint
  {http://arxiv.org/abs/1904.04842} {arXiv:1904.04842 [gr-qc]} \BibitemShut
  {NoStop}%
\bibitem [{\citenamefont {{The LIGO Scientific Collaboration}}\ and\
  \citenamefont {{the Virgo Collaboration}}(2019)}]{GWTC1}%
  \BibitemOpen
  \bibfield  {author} {\bibinfo {author} {\bibnamefont {{The LIGO Scientific
  Collaboration}}}\ and\ \bibinfo {author} {\bibnamefont {{the Virgo
  Collaboration}}},\ }\href {\doibase 10.1103/PhysRevX.9.031040} {\bibfield
  {journal} {\bibinfo  {journal} {Physical Review X}\ }\textbf {\bibinfo
  {volume} {9}},\ \bibinfo {eid} {031040} (\bibinfo {year} {2019})},\ \Eprint
  {http://arxiv.org/abs/1811.12907} {arXiv:1811.12907 [astro-ph.HE]}
  \BibitemShut {NoStop}%
\bibitem [{\citenamefont {{The LIGO Scientific Collaboration}}\ and\
  \citenamefont {{the Virgo Collaboration}}(2020)}]{GWTC2}%
  \BibitemOpen
  \bibfield  {author} {\bibinfo {author} {\bibnamefont {{The LIGO Scientific
  Collaboration}}}\ and\ \bibinfo {author} {\bibnamefont {{the Virgo
  Collaboration}}},\ }\href@noop {} {\bibfield  {journal} {\bibinfo  {journal}
  {arXiv e-prints}\ ,\ \bibinfo {eid} {arXiv:2010.14527}} (\bibinfo {year}
  {2020})},\ \Eprint {http://arxiv.org/abs/2010.14527} {arXiv:2010.14527
  [gr-qc]} \BibitemShut {NoStop}%
\bibitem [{\citenamefont {{Schutz}}(1986)}]{schutz_1986}%
  \BibitemOpen
  \bibfield  {author} {\bibinfo {author} {\bibfnamefont {B.~F.}\ \bibnamefont
  {{Schutz}}},\ }\href {\doibase 10.1038/323310a0} {\bibfield  {journal}
  {\bibinfo  {journal} {\nat}\ }\textbf {\bibinfo {volume} {323}},\ \bibinfo
  {pages} {310} (\bibinfo {year} {1986})}\BibitemShut {NoStop}%
\bibitem [{\citenamefont {{Torres-Orjuela}}\ \emph
  {et~al.}(2020{\natexlab{b}})\citenamefont {{Torres-Orjuela}}, \citenamefont
  {{Chen}},\ and\ \citenamefont {{Amaro-Seoane}}}]{torres-orjuela_chen_2020b}%
  \BibitemOpen
  \bibfield  {author} {\bibinfo {author} {\bibfnamefont {A.}~\bibnamefont
  {{Torres-Orjuela}}}, \bibinfo {author} {\bibfnamefont {X.}~\bibnamefont
  {{Chen}}}, \ and\ \bibinfo {author} {\bibfnamefont {P.}~\bibnamefont
  {{Amaro-Seoane}}},\ }\href@noop {} {\bibfield  {journal} {\bibinfo  {journal}
  {arXiv e-prints}\ ,\ \bibinfo {eid} {arXiv:2010.15856}} (\bibinfo {year}
  {2020}{\natexlab{b}})},\ \Eprint {http://arxiv.org/abs/2010.15856}
  {arXiv:2010.15856 [astro-ph.CO]} \BibitemShut {NoStop}%
\bibitem [{\citenamefont {{Amaro-Seoane}}(2018)}]{amaro-seoane_2018a}%
  \BibitemOpen
  \bibfield  {author} {\bibinfo {author} {\bibfnamefont {P.}~\bibnamefont
  {{Amaro-Seoane}}},\ }\href {\doibase 10.1007/s41114-018-0013-8} {\bibfield
  {journal} {\bibinfo  {journal} {Living Reviews in Relativity}\ }\textbf
  {\bibinfo {volume} {21}},\ \bibinfo {eid} {4} (\bibinfo {year} {2018})},\
  \Eprint {http://arxiv.org/abs/1205.5240} {arXiv:1205.5240 [astro-ph.CO]}
  \BibitemShut {NoStop}%
\bibitem [{\citenamefont {Scrimgeour}\ \emph {et~al.}(2015)\citenamefont
  {Scrimgeour}, \citenamefont {Davis}, \citenamefont {Blake}, \citenamefont
  {Staveley-Smith}, \citenamefont {Magoulas}, \citenamefont {Springob},
  \citenamefont {Beutler}, \citenamefont {Colless}, \citenamefont {Johnson},
  \citenamefont {Jones}, \citenamefont {Koda}, \citenamefont {Lucey},
  \citenamefont {Ma}, \citenamefont {Mould},\ and\ \citenamefont
  {Poole}}]{6dF2015}%
  \BibitemOpen
  \bibfield  {author} {\bibinfo {author} {\bibfnamefont {M.~I.}\ \bibnamefont
  {Scrimgeour}}, \bibinfo {author} {\bibfnamefont {T.~M.}\ \bibnamefont
  {Davis}}, \bibinfo {author} {\bibfnamefont {C.}~\bibnamefont {Blake}},
  \bibinfo {author} {\bibfnamefont {L.}~\bibnamefont {Staveley-Smith}},
  \bibinfo {author} {\bibfnamefont {C.}~\bibnamefont {Magoulas}}, \bibinfo
  {author} {\bibfnamefont {C.~M.}\ \bibnamefont {Springob}}, \bibinfo {author}
  {\bibfnamefont {F.}~\bibnamefont {Beutler}}, \bibinfo {author} {\bibfnamefont
  {M.}~\bibnamefont {Colless}}, \bibinfo {author} {\bibfnamefont
  {A.}~\bibnamefont {Johnson}}, \bibinfo {author} {\bibfnamefont {D.~H.}\
  \bibnamefont {Jones}}, \bibinfo {author} {\bibfnamefont {J.}~\bibnamefont
  {Koda}}, \bibinfo {author} {\bibfnamefont {J.~R.}\ \bibnamefont {Lucey}},
  \bibinfo {author} {\bibfnamefont {Y.-Z.}\ \bibnamefont {Ma}}, \bibinfo
  {author} {\bibfnamefont {J.}~\bibnamefont {Mould}}, \ and\ \bibinfo {author}
  {\bibfnamefont {G.~B.}\ \bibnamefont {Poole}},\ }\href {\doibase
  10.1093/mnras/stv2146} {\bibfield  {journal} {\bibinfo  {journal} {Monthly
  Notices of the Royal Astronomical Society}\ }\textbf {\bibinfo {volume}
  {455}},\ \bibinfo {pages} {386} (\bibinfo {year} {2015})},\ \Eprint
  {http://arxiv.org/abs/https://academic.oup.com/mnras/article-pdf/455/1/386/3059232/stv2146.pdf}
  {https://academic.oup.com/mnras/article-pdf/455/1/386/3059232/stv2146.pdf}
  \BibitemShut {NoStop}%
\bibitem [{\citenamefont {{Colin}}\ \emph {et~al.}(2017)\citenamefont
  {{Colin}}, \citenamefont {{Mohayaee}}, \citenamefont {{Rameez}},\ and\
  \citenamefont {{Sarkar}}}]{ColinEtAl2017}%
  \BibitemOpen
  \bibfield  {author} {\bibinfo {author} {\bibfnamefont {J.}~\bibnamefont
  {{Colin}}}, \bibinfo {author} {\bibfnamefont {R.}~\bibnamefont {{Mohayaee}}},
  \bibinfo {author} {\bibfnamefont {M.}~\bibnamefont {{Rameez}}}, \ and\
  \bibinfo {author} {\bibfnamefont {S.}~\bibnamefont {{Sarkar}}},\ }\href
  {\doibase 10.1093/mnras/stx1631} {\bibfield  {journal} {\bibinfo  {journal}
  {MNRAS}\ }\textbf {\bibinfo {volume} {471}},\ \bibinfo {pages} {1045}
  (\bibinfo {year} {2017})},\ \Eprint {http://arxiv.org/abs/1703.09376}
  {arXiv:1703.09376 [astro-ph.CO]} \BibitemShut {NoStop}%
\bibitem [{\citenamefont {{Bahcall}}(1988)}]{Bahcall1988}%
  \BibitemOpen
  \bibfield  {author} {\bibinfo {author} {\bibfnamefont {N.~A.}\ \bibnamefont
  {{Bahcall}}},\ }\href {\doibase 10.1146/annurev.aa.26.090188.003215}
  {\bibfield  {journal} {\bibinfo  {journal} {ARA\&A}\ }\textbf {\bibinfo
  {volume} {26}},\ \bibinfo {pages} {631} (\bibinfo {year} {1988})}\BibitemShut
  {NoStop}%
\bibitem [{\citenamefont {{Zwicky}}(1933)}]{Zwicky1933}%
  \BibitemOpen
  \bibfield  {author} {\bibinfo {author} {\bibfnamefont {F.}~\bibnamefont
  {{Zwicky}}},\ }\href@noop {} {\bibfield  {journal} {\bibinfo  {journal}
  {Helvetica Physica Acta}\ }\textbf {\bibinfo {volume} {6}},\ \bibinfo {pages}
  {110} (\bibinfo {year} {1933})}\BibitemShut {NoStop}%
\bibitem [{\citenamefont {{Girardi}}\ \emph {et~al.}(1996)\citenamefont
  {{Girardi}}, \citenamefont {{Fadda}}, \citenamefont {{Giuricin}},
  \citenamefont {{Mardirossian}}, \citenamefont {{Mezzetti}},\ and\
  \citenamefont {{Biviano}}}]{GirardiEtAl1996}%
  \BibitemOpen
  \bibfield  {author} {\bibinfo {author} {\bibfnamefont {M.}~\bibnamefont
  {{Girardi}}}, \bibinfo {author} {\bibfnamefont {D.}~\bibnamefont {{Fadda}}},
  \bibinfo {author} {\bibfnamefont {G.}~\bibnamefont {{Giuricin}}}, \bibinfo
  {author} {\bibfnamefont {F.}~\bibnamefont {{Mardirossian}}}, \bibinfo
  {author} {\bibfnamefont {M.}~\bibnamefont {{Mezzetti}}}, \ and\ \bibinfo
  {author} {\bibfnamefont {A.}~\bibnamefont {{Biviano}}},\ }\href {\doibase
  10.1086/176711} {\bibfield  {journal} {\bibinfo  {journal} {ApJ}\ }\textbf
  {\bibinfo {volume} {457}},\ \bibinfo {pages} {61} (\bibinfo {year} {1996})},\
  \Eprint {http://arxiv.org/abs/astro-ph/9507031} {arXiv:astro-ph/9507031
  [astro-ph]} \BibitemShut {NoStop}%
\bibitem [{\citenamefont {{Ruel}}\ \emph {et~al.}(2014)\citenamefont {{Ruel}},
  \citenamefont {{Bazin}}, \citenamefont {{Bayliss}}, \citenamefont
  {{Brodwin}}, \citenamefont {{Foley}}, \citenamefont {{Stalder}},
  \citenamefont {{Aird}}, \citenamefont {{Armstrong}}, \citenamefont {{Ashby}},
  \citenamefont {{Bautz}}, \citenamefont {{Benson}}, \citenamefont {{Bleem}},
  \citenamefont {{Bocquet}}, \citenamefont {{Carlstrom}}, \citenamefont
  {{Chang}}, \citenamefont {{Chapman}}, \citenamefont {{Cho}}, \citenamefont
  {{Clocchiatti}}, \citenamefont {{Crawford}}, \citenamefont {{Crites}},
  \citenamefont {{de Haan}}, \citenamefont {{Desai}}, \citenamefont {{Dobbs}},
  \citenamefont {{Dudley}}, \citenamefont {{Forman}}, \citenamefont {{George}},
  \citenamefont {{Gladders}}, \citenamefont {{Gonzalez}}, \citenamefont
  {{Halverson}}, \citenamefont {{Harrington}}, \citenamefont {{High}},
  \citenamefont {{Holder}}, \citenamefont {{Holzapfel}}, \citenamefont
  {{Hrubes}}, \citenamefont {{Jones}}, \citenamefont {{Joy}}, \citenamefont
  {{Keisler}}, \citenamefont {{Knox}}, \citenamefont {{Lee}}, \citenamefont
  {{Leitch}}, \citenamefont {{Liu}}, \citenamefont {{Lueker}}, \citenamefont
  {{Luong-Van}}, \citenamefont {{Mantz}}, \citenamefont {{Marrone}},
  \citenamefont {{McDonald}}, \citenamefont {{McMahon}}, \citenamefont
  {{Mehl}}, \citenamefont {{Meyer}}, \citenamefont {{Mocanu}}, \citenamefont
  {{Mohr}}, \citenamefont {{Montroy}}, \citenamefont {{Murray}}, \citenamefont
  {{Natoli}}, \citenamefont {{Nurgaliev}}, \citenamefont {{Padin}},
  \citenamefont {{Plagge}}, \citenamefont {{Pryke}}, \citenamefont
  {{Reichardt}}, \citenamefont {{Rest}}, \citenamefont {{Ruhl}}, \citenamefont
  {{Saliwanchik}}, \citenamefont {{Saro}}, \citenamefont {{Sayre}},
  \citenamefont {{Schaffer}}, \citenamefont {{Shaw}}, \citenamefont
  {{Shirokoff}}, \citenamefont {{Song}}, \citenamefont {{{\v{S}}uhada}},
  \citenamefont {{Spieler}}, \citenamefont {{Stanford}}, \citenamefont
  {{Staniszewski}}, \citenamefont {{Starsk}}, \citenamefont {{Story}},
  \citenamefont {{Stubbs}}, \citenamefont {{van Engelen}}, \citenamefont
  {{Vanderlinde}}, \citenamefont {{Vieira}}, \citenamefont {{Vikhlinin}},
  \citenamefont {{Williamson}}, \citenamefont {{Zahn}},\ and\ \citenamefont
  {{Zenteno}}}]{RuelEtAl2014}%
  \BibitemOpen
  \bibfield  {author} {\bibinfo {author} {\bibfnamefont {J.}~\bibnamefont
  {{Ruel}}}, \bibinfo {author} {\bibfnamefont {G.}~\bibnamefont {{Bazin}}},
  \bibinfo {author} {\bibfnamefont {M.}~\bibnamefont {{Bayliss}}}, \bibinfo
  {author} {\bibfnamefont {M.}~\bibnamefont {{Brodwin}}}, \bibinfo {author}
  {\bibfnamefont {R.~J.}\ \bibnamefont {{Foley}}}, \bibinfo {author}
  {\bibfnamefont {B.}~\bibnamefont {{Stalder}}}, \bibinfo {author}
  {\bibfnamefont {K.~A.}\ \bibnamefont {{Aird}}}, \bibinfo {author}
  {\bibfnamefont {R.}~\bibnamefont {{Armstrong}}}, \bibinfo {author}
  {\bibfnamefont {M.~L.~N.}\ \bibnamefont {{Ashby}}}, \bibinfo {author}
  {\bibfnamefont {M.}~\bibnamefont {{Bautz}}}, \bibinfo {author} {\bibfnamefont
  {B.~A.}\ \bibnamefont {{Benson}}}, \bibinfo {author} {\bibfnamefont {L.~E.}\
  \bibnamefont {{Bleem}}}, \bibinfo {author} {\bibfnamefont {S.}~\bibnamefont
  {{Bocquet}}}, \bibinfo {author} {\bibfnamefont {J.~E.}\ \bibnamefont
  {{Carlstrom}}}, \bibinfo {author} {\bibfnamefont {C.~L.}\ \bibnamefont
  {{Chang}}}, \bibinfo {author} {\bibfnamefont {S.~C.}\ \bibnamefont
  {{Chapman}}}, \bibinfo {author} {\bibfnamefont {H.~M.}\ \bibnamefont
  {{Cho}}}, \bibinfo {author} {\bibfnamefont {A.}~\bibnamefont
  {{Clocchiatti}}}, \bibinfo {author} {\bibfnamefont {T.~M.}\ \bibnamefont
  {{Crawford}}}, \bibinfo {author} {\bibfnamefont {A.~T.}\ \bibnamefont
  {{Crites}}}, \bibinfo {author} {\bibfnamefont {T.}~\bibnamefont {{de Haan}}},
  \bibinfo {author} {\bibfnamefont {S.}~\bibnamefont {{Desai}}}, \bibinfo
  {author} {\bibfnamefont {M.~A.}\ \bibnamefont {{Dobbs}}}, \bibinfo {author}
  {\bibfnamefont {J.~P.}\ \bibnamefont {{Dudley}}}, \bibinfo {author}
  {\bibfnamefont {W.~R.}\ \bibnamefont {{Forman}}}, \bibinfo {author}
  {\bibfnamefont {E.~M.}\ \bibnamefont {{George}}}, \bibinfo {author}
  {\bibfnamefont {M.~D.}\ \bibnamefont {{Gladders}}}, \bibinfo {author}
  {\bibfnamefont {A.~H.}\ \bibnamefont {{Gonzalez}}}, \bibinfo {author}
  {\bibfnamefont {N.~W.}\ \bibnamefont {{Halverson}}}, \bibinfo {author}
  {\bibfnamefont {N.~L.}\ \bibnamefont {{Harrington}}}, \bibinfo {author}
  {\bibfnamefont {F.~W.}\ \bibnamefont {{High}}}, \bibinfo {author}
  {\bibfnamefont {G.~P.}\ \bibnamefont {{Holder}}}, \bibinfo {author}
  {\bibfnamefont {W.~L.}\ \bibnamefont {{Holzapfel}}}, \bibinfo {author}
  {\bibfnamefont {J.~D.}\ \bibnamefont {{Hrubes}}}, \bibinfo {author}
  {\bibfnamefont {C.}~\bibnamefont {{Jones}}}, \bibinfo {author} {\bibfnamefont
  {M.}~\bibnamefont {{Joy}}}, \bibinfo {author} {\bibfnamefont
  {R.}~\bibnamefont {{Keisler}}}, \bibinfo {author} {\bibfnamefont
  {L.}~\bibnamefont {{Knox}}}, \bibinfo {author} {\bibfnamefont {A.~T.}\
  \bibnamefont {{Lee}}}, \bibinfo {author} {\bibfnamefont {E.~M.}\ \bibnamefont
  {{Leitch}}}, \bibinfo {author} {\bibfnamefont {J.}~\bibnamefont {{Liu}}},
  \bibinfo {author} {\bibfnamefont {M.}~\bibnamefont {{Lueker}}}, \bibinfo
  {author} {\bibfnamefont {D.}~\bibnamefont {{Luong-Van}}}, \bibinfo {author}
  {\bibfnamefont {A.}~\bibnamefont {{Mantz}}}, \bibinfo {author} {\bibfnamefont
  {D.~P.}\ \bibnamefont {{Marrone}}}, \bibinfo {author} {\bibfnamefont
  {M.}~\bibnamefont {{McDonald}}}, \bibinfo {author} {\bibfnamefont {J.~J.}\
  \bibnamefont {{McMahon}}}, \bibinfo {author} {\bibfnamefont {J.}~\bibnamefont
  {{Mehl}}}, \bibinfo {author} {\bibfnamefont {S.~S.}\ \bibnamefont {{Meyer}}},
  \bibinfo {author} {\bibfnamefont {L.}~\bibnamefont {{Mocanu}}}, \bibinfo
  {author} {\bibfnamefont {J.~J.}\ \bibnamefont {{Mohr}}}, \bibinfo {author}
  {\bibfnamefont {T.~E.}\ \bibnamefont {{Montroy}}}, \bibinfo {author}
  {\bibfnamefont {S.~S.}\ \bibnamefont {{Murray}}}, \bibinfo {author}
  {\bibfnamefont {T.}~\bibnamefont {{Natoli}}}, \bibinfo {author}
  {\bibfnamefont {D.}~\bibnamefont {{Nurgaliev}}}, \bibinfo {author}
  {\bibfnamefont {S.}~\bibnamefont {{Padin}}}, \bibinfo {author} {\bibfnamefont
  {T.}~\bibnamefont {{Plagge}}}, \bibinfo {author} {\bibfnamefont
  {C.}~\bibnamefont {{Pryke}}}, \bibinfo {author} {\bibfnamefont {C.~L.}\
  \bibnamefont {{Reichardt}}}, \bibinfo {author} {\bibfnamefont
  {A.}~\bibnamefont {{Rest}}}, \bibinfo {author} {\bibfnamefont {J.~E.}\
  \bibnamefont {{Ruhl}}}, \bibinfo {author} {\bibfnamefont {B.~R.}\
  \bibnamefont {{Saliwanchik}}}, \bibinfo {author} {\bibfnamefont
  {A.}~\bibnamefont {{Saro}}}, \bibinfo {author} {\bibfnamefont {J.~T.}\
  \bibnamefont {{Sayre}}}, \bibinfo {author} {\bibfnamefont {K.~K.}\
  \bibnamefont {{Schaffer}}}, \bibinfo {author} {\bibfnamefont
  {L.}~\bibnamefont {{Shaw}}}, \bibinfo {author} {\bibfnamefont
  {E.}~\bibnamefont {{Shirokoff}}}, \bibinfo {author} {\bibfnamefont
  {J.}~\bibnamefont {{Song}}}, \bibinfo {author} {\bibfnamefont
  {R.}~\bibnamefont {{{\v{S}}uhada}}}, \bibinfo {author} {\bibfnamefont
  {H.~G.}\ \bibnamefont {{Spieler}}}, \bibinfo {author} {\bibfnamefont {S.~A.}\
  \bibnamefont {{Stanford}}}, \bibinfo {author} {\bibfnamefont
  {Z.}~\bibnamefont {{Staniszewski}}}, \bibinfo {author} {\bibfnamefont
  {A.~A.}\ \bibnamefont {{Starsk}}}, \bibinfo {author} {\bibfnamefont
  {K.}~\bibnamefont {{Story}}}, \bibinfo {author} {\bibfnamefont {C.~W.}\
  \bibnamefont {{Stubbs}}}, \bibinfo {author} {\bibfnamefont {A.}~\bibnamefont
  {{van Engelen}}}, \bibinfo {author} {\bibfnamefont {K.}~\bibnamefont
  {{Vanderlinde}}}, \bibinfo {author} {\bibfnamefont {J.~D.}\ \bibnamefont
  {{Vieira}}}, \bibinfo {author} {\bibfnamefont {A.}~\bibnamefont
  {{Vikhlinin}}}, \bibinfo {author} {\bibfnamefont {R.}~\bibnamefont
  {{Williamson}}}, \bibinfo {author} {\bibfnamefont {O.}~\bibnamefont
  {{Zahn}}}, \ and\ \bibinfo {author} {\bibfnamefont {A.}~\bibnamefont
  {{Zenteno}}},\ }\href {\doibase 10.1088/0004-637X/792/1/45} {\bibfield
  {journal} {\bibinfo  {journal} {ApJ}\ }\textbf {\bibinfo {volume} {792}},\
  \bibinfo {eid} {45} (\bibinfo {year} {2014})},\ \Eprint
  {http://arxiv.org/abs/1311.4953} {arXiv:1311.4953 [astro-ph.CO]} \BibitemShut
  {NoStop}%
\bibitem [{\citenamefont {Johnson}\ and\ \citenamefont
  {Teller}(1982)}]{johnson_teller_1982}%
  \BibitemOpen
  \bibfield  {author} {\bibinfo {author} {\bibfnamefont {M.~H.}\ \bibnamefont
  {Johnson}}\ and\ \bibinfo {author} {\bibfnamefont {E.}~\bibnamefont
  {Teller}},\ }\href {\doibase 10.1073/pnas.79.4.1340} {\bibfield  {journal}
  {\bibinfo  {journal} {Proceedings of the National Academy of Sciences}\
  }\textbf {\bibinfo {volume} {79}},\ \bibinfo {pages} {1340} (\bibinfo {year}
  {1982})},\ \Eprint
  {http://arxiv.org/abs/https://www.pnas.org/content/79/4/1340.full.pdf}
  {https://www.pnas.org/content/79/4/1340.full.pdf} \BibitemShut {NoStop}%
\bibitem [{\citenamefont {Thorne}(1980)}]{thorne_1980}%
  \BibitemOpen
  \bibfield  {author} {\bibinfo {author} {\bibfnamefont {K.~S.}\ \bibnamefont
  {Thorne}},\ }\href {\doibase 10.1103/RevModPhys.52.299} {\bibfield  {journal}
  {\bibinfo  {journal} {Rev. Mod. Phys.}\ }\textbf {\bibinfo {volume} {52}},\
  \bibinfo {pages} {299} (\bibinfo {year} {1980})}\BibitemShut {NoStop}%
\bibitem [{\citenamefont {{Ruiz}}\ \emph {et~al.}(2008)\citenamefont {{Ruiz}},
  \citenamefont {{Alcubierre}}, \citenamefont {{N{\'u}{\~n}ez}},\ and\
  \citenamefont {{Takahashi}}}]{ruiz_alcubierre_2008}%
  \BibitemOpen
  \bibfield  {author} {\bibinfo {author} {\bibfnamefont {M.}~\bibnamefont
  {{Ruiz}}}, \bibinfo {author} {\bibfnamefont {M.}~\bibnamefont
  {{Alcubierre}}}, \bibinfo {author} {\bibfnamefont {D.}~\bibnamefont
  {{N{\'u}{\~n}ez}}}, \ and\ \bibinfo {author} {\bibfnamefont {R.}~\bibnamefont
  {{Takahashi}}},\ }\href {\doibase 10.1007/s10714-008-0684-7} {\bibfield
  {journal} {\bibinfo  {journal} {General Relativity and Gravitation}\ }\textbf
  {\bibinfo {volume} {40}},\ \bibinfo {pages} {2467} (\bibinfo {year}
  {2008})}\BibitemShut {NoStop}%
\bibitem [{\citenamefont {{Goldberg}}\ \emph {et~al.}(1967)\citenamefont
  {{Goldberg}}, \citenamefont {{Macfarlane}}, \citenamefont {{Newman}},
  \citenamefont {{Rohrlich}},\ and\ \citenamefont
  {{Sudarshan}}}]{goldberg_macfarlane_1967}%
  \BibitemOpen
  \bibfield  {author} {\bibinfo {author} {\bibfnamefont {J.~N.}\ \bibnamefont
  {{Goldberg}}}, \bibinfo {author} {\bibfnamefont {A.~J.}\ \bibnamefont
  {{Macfarlane}}}, \bibinfo {author} {\bibfnamefont {E.~T.}\ \bibnamefont
  {{Newman}}}, \bibinfo {author} {\bibfnamefont {F.}~\bibnamefont
  {{Rohrlich}}}, \ and\ \bibinfo {author} {\bibfnamefont {E.~C.~G.}\
  \bibnamefont {{Sudarshan}}},\ }\href {\doibase 10.1063/1.1705135} {\bibfield
  {journal} {\bibinfo  {journal} {Journal of Mathematical Physics}\ }\textbf
  {\bibinfo {volume} {8}},\ \bibinfo {pages} {2155} (\bibinfo {year}
  {1967})}\BibitemShut {NoStop}%
\bibitem [{\citenamefont {{Arun}}\ \emph {et~al.}(2009)\citenamefont {{Arun}},
  \citenamefont {{Buonanno}}, \citenamefont {{Faye}},\ and\ \citenamefont
  {{Ochsner}}}]{ArunEtAl2009}%
  \BibitemOpen
  \bibfield  {author} {\bibinfo {author} {\bibfnamefont {K.~G.}\ \bibnamefont
  {{Arun}}}, \bibinfo {author} {\bibfnamefont {A.}~\bibnamefont {{Buonanno}}},
  \bibinfo {author} {\bibfnamefont {G.}~\bibnamefont {{Faye}}}, \ and\ \bibinfo
  {author} {\bibfnamefont {E.}~\bibnamefont {{Ochsner}}},\ }\href {\doibase
  10.1103/PhysRevD.79.104023} {\bibfield  {journal} {\bibinfo  {journal} {Phys.
  Rev. D.}\ }\textbf {\bibinfo {volume} {79}},\ \bibinfo {eid} {104023}
  (\bibinfo {year} {2009})},\ \Eprint {http://arxiv.org/abs/0810.5336}
  {arXiv:0810.5336 [gr-qc]} \BibitemShut {NoStop}%
\bibitem [{\citenamefont {{Helstrom}}(1968)}]{Helstrom68}%
  \BibitemOpen
  \bibfield  {author} {\bibinfo {author} {\bibfnamefont {C.~W.}\ \bibnamefont
  {{Helstrom}}},\ }\href@noop {} {\emph {\bibinfo {title} {{Statistical Theory
  of Signal Detection}}}},\ edited by\ \bibinfo {editor} {\bibfnamefont
  {L.}~\bibnamefont {Pergamon~Press}}\ (\bibinfo {year} {1968})\BibitemShut
  {NoStop}%
\bibitem [{\citenamefont {{Thorne}}(1987)}]{Thorne87}%
  \BibitemOpen
  \bibfield  {author} {\bibinfo {author} {\bibfnamefont {K.~S.}\ \bibnamefont
  {{Thorne}}},\ }\enquote {\bibinfo {title} {{Gravitational radiation.}}}\ in\
  \href@noop {} {\emph {\bibinfo {booktitle} {Three hundred years of
  gravitation, p.~330 - 458}}},\ \bibinfo {editor} {edited by\ \bibinfo
  {editor} {\bibfnamefont {S.~W.}\ \bibnamefont {Hawking}}\ and\ \bibinfo
  {editor} {\bibfnamefont {W.}~\bibnamefont {Israel}}}\ (\bibinfo {year}
  {1987})\ pp.\ \bibinfo {pages} {330--458}\BibitemShut {NoStop}%
\bibitem [{\citenamefont {{Finn}}(1992)}]{Finn92}%
  \BibitemOpen
  \bibfield  {author} {\bibinfo {author} {\bibfnamefont {L.~S.}\ \bibnamefont
  {{Finn}}},\ }\href {\doibase 10.1103/PhysRevD.46.5236} {\bibfield  {journal}
  {\bibinfo  {journal} {prd}\ }\textbf {\bibinfo {volume} {46}},\ \bibinfo
  {pages} {5236} (\bibinfo {year} {1992})},\ \Eprint
  {http://arxiv.org/abs/arXiv:gr-qc/9209010} {arXiv:gr-qc/9209010} \BibitemShut
  {NoStop}%
\bibitem [{\citenamefont {{Klein}}\ \emph {et~al.}(2016)\citenamefont
  {{Klein}}, \citenamefont {{Barausse}}, \citenamefont {{Sesana}},
  \citenamefont {{Petiteau}}, \citenamefont {{Berti}}, \citenamefont {{Babak}},
  \citenamefont {{Gair}}, \citenamefont {{Aoudia}}, \citenamefont {{Hinder}},
  \citenamefont {{Ohme}},\ and\ \citenamefont
  {{Wardell}}}]{2016PhRvD..93b4003K}%
  \BibitemOpen
  \bibfield  {author} {\bibinfo {author} {\bibfnamefont {A.}~\bibnamefont
  {{Klein}}}, \bibinfo {author} {\bibfnamefont {E.}~\bibnamefont {{Barausse}}},
  \bibinfo {author} {\bibfnamefont {A.}~\bibnamefont {{Sesana}}}, \bibinfo
  {author} {\bibfnamefont {A.}~\bibnamefont {{Petiteau}}}, \bibinfo {author}
  {\bibfnamefont {E.}~\bibnamefont {{Berti}}}, \bibinfo {author} {\bibfnamefont
  {S.}~\bibnamefont {{Babak}}}, \bibinfo {author} {\bibfnamefont
  {J.}~\bibnamefont {{Gair}}}, \bibinfo {author} {\bibfnamefont
  {S.}~\bibnamefont {{Aoudia}}}, \bibinfo {author} {\bibfnamefont
  {I.}~\bibnamefont {{Hinder}}}, \bibinfo {author} {\bibfnamefont
  {F.}~\bibnamefont {{Ohme}}}, \ and\ \bibinfo {author} {\bibfnamefont
  {B.}~\bibnamefont {{Wardell}}},\ }\href {\doibase 10.1103/PhysRevD.93.024003}
  {\bibfield  {journal} {\bibinfo  {journal} {Ph. Rv. D}\ }\textbf {\bibinfo
  {volume} {93}},\ \bibinfo {eid} {024003} (\bibinfo {year} {2016})},\ \Eprint
  {http://arxiv.org/abs/1511.05581} {arXiv:1511.05581 [gr-qc]} \BibitemShut
  {NoStop}%
\bibitem [{\citenamefont {{Cutler}}\ and\ \citenamefont
  {{Vallisneri}}(2007)}]{CutlerVallisneri2007}%
  \BibitemOpen
  \bibfield  {author} {\bibinfo {author} {\bibfnamefont {C.}~\bibnamefont
  {{Cutler}}}\ and\ \bibinfo {author} {\bibfnamefont {M.}~\bibnamefont
  {{Vallisneri}}},\ }\href {\doibase 10.1103/PhysRevD.76.104018} {\bibfield
  {journal} {\bibinfo  {journal} {Ph. Rv. D.}\ }\textbf {\bibinfo {volume}
  {76}},\ \bibinfo {eid} {104018} (\bibinfo {year} {2007})},\ \Eprint
  {http://arxiv.org/abs/0707.2982} {arXiv:0707.2982 [gr-qc]} \BibitemShut
  {NoStop}%
\bibitem [{\citenamefont {{Coe}}(2009)}]{coe_2009}%
  \BibitemOpen
  \bibfield  {author} {\bibinfo {author} {\bibfnamefont {D.}~\bibnamefont
  {{Coe}}},\ }\href@noop {} {\bibfield  {journal} {\bibinfo  {journal} {arXiv
  e-prints}\ ,\ \bibinfo {eid} {arXiv:0906.4123}} (\bibinfo {year} {2009})},\
  \Eprint {http://arxiv.org/abs/0906.4123} {arXiv:0906.4123 [astro-ph.IM]}
  \BibitemShut {NoStop}%
\bibitem [{\citenamefont {{Chua}}\ \emph {et~al.}(2020)\citenamefont {{Chua}},
  \citenamefont {{Katz}}, \citenamefont {{Warburton}},\ and\ \citenamefont
  {{Hughes}}}]{ChuaEtAl2020}%
  \BibitemOpen
  \bibfield  {author} {\bibinfo {author} {\bibfnamefont {A.~J.~K.}\
  \bibnamefont {{Chua}}}, \bibinfo {author} {\bibfnamefont {M.~L.}\
  \bibnamefont {{Katz}}}, \bibinfo {author} {\bibfnamefont {N.}~\bibnamefont
  {{Warburton}}}, \ and\ \bibinfo {author} {\bibfnamefont {S.~A.}\ \bibnamefont
  {{Hughes}}},\ }\href@noop {} {\bibfield  {journal} {\bibinfo  {journal}
  {arXiv e-prints}\ ,\ \bibinfo {eid} {arXiv:2008.06071}} (\bibinfo {year}
  {2020})},\ \Eprint {http://arxiv.org/abs/2008.06071} {arXiv:2008.06071
  [gr-qc]} \BibitemShut {NoStop}%
\bibitem [{\citenamefont {Peters}\ and\ \citenamefont
  {Mathews}(1963)}]{peters_mathews_1963}%
  \BibitemOpen
  \bibfield  {author} {\bibinfo {author} {\bibfnamefont {P.}~\bibnamefont
  {Peters}}\ and\ \bibinfo {author} {\bibfnamefont {J.}~\bibnamefont
  {Mathews}},\ }\href {\doibase 10.1103/PhysRev.131.435} {\bibfield  {journal}
  {\bibinfo  {journal} {Physical Review (U.S.) Superseded in part by Phys. Rev.
  A, Phys. Rev. B: Solid State, Phys. Rev. C, and Phys. Rev. D}\ }\textbf
  {\bibinfo {volume} {131}} (\bibinfo {year} {1963}),\
  10.1103/PhysRev.131.435}\BibitemShut {NoStop}%
\bibitem [{\citenamefont {{Babak}}\ \emph {et~al.}(2017)\citenamefont
  {{Babak}}, \citenamefont {{Gair}}, \citenamefont {{Sesana}}, \citenamefont
  {{Barausse}}, \citenamefont {{Sopuerta}}, \citenamefont {{Berry}},
  \citenamefont {{Berti}}, \citenamefont {{Amaro-Seoane}}, \citenamefont
  {{Petiteau}},\ and\ \citenamefont {{Klein}}}]{BabakEtAl2017}%
  \BibitemOpen
  \bibfield  {author} {\bibinfo {author} {\bibfnamefont {S.}~\bibnamefont
  {{Babak}}}, \bibinfo {author} {\bibfnamefont {J.}~\bibnamefont {{Gair}}},
  \bibinfo {author} {\bibfnamefont {A.}~\bibnamefont {{Sesana}}}, \bibinfo
  {author} {\bibfnamefont {E.}~\bibnamefont {{Barausse}}}, \bibinfo {author}
  {\bibfnamefont {C.~F.}\ \bibnamefont {{Sopuerta}}}, \bibinfo {author}
  {\bibfnamefont {C.~P.~L.}\ \bibnamefont {{Berry}}}, \bibinfo {author}
  {\bibfnamefont {E.}~\bibnamefont {{Berti}}}, \bibinfo {author} {\bibfnamefont
  {P.}~\bibnamefont {{Amaro-Seoane}}}, \bibinfo {author} {\bibfnamefont
  {A.}~\bibnamefont {{Petiteau}}}, \ and\ \bibinfo {author} {\bibfnamefont
  {A.}~\bibnamefont {{Klein}}},\ }\href {\doibase 10.1103/PhysRevD.95.103012}
  {\bibfield  {journal} {\bibinfo  {journal} {Phys. Rev. D}\ }\textbf {\bibinfo
  {volume} {95}},\ \bibinfo {eid} {103012} (\bibinfo {year} {2017})},\ \Eprint
  {http://arxiv.org/abs/1703.09722} {arXiv:1703.09722 [gr-qc]} \BibitemShut
  {NoStop}%
\bibitem [{\citenamefont {{McKernan}}\ \emph {et~al.}(2019)\citenamefont
  {{McKernan}}, \citenamefont {{Ford}},\ and\ \citenamefont
  {{Larson}}}]{mckernan_ford_2019}%
  \BibitemOpen
  \bibfield  {author} {\bibinfo {author} {\bibfnamefont {B.}~\bibnamefont
  {{McKernan}}}, \bibinfo {author} {\bibfnamefont {K.~E.~S.}\ \bibnamefont
  {{Ford}}}, \ and\ \bibinfo {author} {\bibfnamefont {S.~L.}\ \bibnamefont
  {{Larson}}},\ }in\ \href@noop {} {\emph {\bibinfo {booktitle} {American
  Astronomical Society Meeting Abstracts \#233}}},\ \bibinfo {series} {American
  Astronomical Society Meeting Abstracts}, Vol.\ \bibinfo {volume} {233}\
  (\bibinfo {year} {2019})\ p.\ \bibinfo {pages} {141.02}\BibitemShut {NoStop}%
\bibitem [{\citenamefont {{Pan}}\ and\ \citenamefont
  {{Yang}}(2021)}]{pan_yang_2021}%
  \BibitemOpen
  \bibfield  {author} {\bibinfo {author} {\bibfnamefont {Z.}~\bibnamefont
  {{Pan}}}\ and\ \bibinfo {author} {\bibfnamefont {H.}~\bibnamefont {{Yang}}},\
  }\href {\doibase 10.1103/PhysRevD.103.103018} {\bibfield  {journal} {\bibinfo
   {journal} {\prd}\ }\textbf {\bibinfo {volume} {103}},\ \bibinfo {eid}
  {103018} (\bibinfo {year} {2021})},\ \Eprint
  {http://arxiv.org/abs/2101.09146} {arXiv:2101.09146 [astro-ph.HE]}
  \BibitemShut {NoStop}%
\bibitem [{\citenamefont {{Pan}}\ \emph {et~al.}(2021)\citenamefont {{Pan}},
  \citenamefont {{Lyu}},\ and\ \citenamefont {{Yang}}}]{pan_lyu_2021}%
  \BibitemOpen
  \bibfield  {author} {\bibinfo {author} {\bibfnamefont {Z.}~\bibnamefont
  {{Pan}}}, \bibinfo {author} {\bibfnamefont {Z.}~\bibnamefont {{Lyu}}}, \ and\
  \bibinfo {author} {\bibfnamefont {H.}~\bibnamefont {{Yang}}},\ }\href@noop {}
  {\bibfield  {journal} {\bibinfo  {journal} {arXiv e-prints}\ ,\ \bibinfo
  {eid} {arXiv:2104.01208}} (\bibinfo {year} {2021})},\ \Eprint
  {http://arxiv.org/abs/2104.01208} {arXiv:2104.01208 [astro-ph.HE]}
  \BibitemShut {NoStop}%
\bibitem [{\citenamefont {{Barausse}}\ and\ \citenamefont
  {{Rezzolla}}(2008)}]{barausse_rezzolla_2008}%
  \BibitemOpen
  \bibfield  {author} {\bibinfo {author} {\bibfnamefont {E.}~\bibnamefont
  {{Barausse}}}\ and\ \bibinfo {author} {\bibfnamefont {L.}~\bibnamefont
  {{Rezzolla}}},\ }\href {\doibase 10.1103/PhysRevD.77.104027} {\bibfield
  {journal} {\bibinfo  {journal} {\prd}\ }\textbf {\bibinfo {volume} {77}},\
  \bibinfo {eid} {104027} (\bibinfo {year} {2008})},\ \Eprint
  {http://arxiv.org/abs/0711.4558} {arXiv:0711.4558 [gr-qc]} \BibitemShut
  {NoStop}%
\bibitem [{\citenamefont {{Amaro-Seoane }}\ and\ \citenamefont {{et
  al.}}(2017)}]{lisa_2017}%
  \BibitemOpen
  \bibfield  {author} {\bibinfo {author} {\bibfnamefont {P.}~\bibnamefont
  {{Amaro-Seoane }}}\ and\ \bibinfo {author} {\bibnamefont {{et al.}}},\
  }\href@noop {} {\bibfield  {journal} {\bibinfo  {journal} {ArXiv e-prints}\ }
  (\bibinfo {year} {2017})},\ \Eprint {http://arxiv.org/abs/1702.00786}
  {arXiv:1702.00786 [astro-ph.IM]} \BibitemShut {NoStop}%
\end{thebibliography}
\end{document}